\documentclass{article}

\usepackage{PRIMEarxiv}

\usepackage[utf8]{inputenc} % allow utf-8 input
\usepackage[T1]{fontenc}    % use 8-bit T1 fonts
\usepackage{hyperref}       % hyperlinks
\usepackage{url}            % simple URL typesetting
\usepackage{booktabs}       % professional-quality tables
\usepackage{amsfonts}       % blackboard math symbols
\usepackage{nicefrac}       % compact symbols for 1/2, etc.
\usepackage{microtype}      % microtypography
\usepackage{lipsum}
\usepackage{fancyhdr}       % header
\usepackage{graphicx}       % graphics
\graphicspath{{media/}}     % organize your images and other figures under media/ folder
\usepackage{float}
\usepackage{tabularx}

\usepackage[numbers,sort&compress]{natbib}

\title{Evaluating the Impact of COVID-19 on Transportation Infrastructure Funding
}

\author{
  Lu Gao, Yunpeng Zhang \\
  University of Houston \\
  Houston, TX \\
  \texttt{\{lgao5, yzhan226\}@central.uh.edu} \\
  \And
  Pan Lu, Yihao Ren \\
  North Dakota State University \\
  Fargo, ND \\
  \texttt{Pan.Lu@ndsu.edu, renren071@gmail.com} \\
  \And
  Fengxiang Qiao \\
  Texas Southern University \\
  Houston, TX \\
  \texttt{fengxiang.qiao@tsu.edu} \\
  \And
  Joshua Qiang Li \\
  Oklahoma State University \\
  Stillwater, OK \\
  \texttt{qiang.li@okstate.edu} \\
}

\begin{document}
\maketitle

\begin{abstract}
The coronavirus disease 2019 (COVID-19) pandemic has caused a reduction in business and routine activity and resulted in less motor fuel consumption. Thus, the gas tax revenue is reduced which is the major funding resource supporting the rehabilitation and maintenance of transportation infrastructure systems. The focus of this study is to evaluate the impact of the COVID-19 pandemic on transportation infrastructure funds in the United States through analyzing the motor fuel consumption data. Machine learning models were developed by integrating COVID-19 scenarios, fuel consumptions, and demographic data. The best model achieves an R2-score of more than 95\% and captures the fluctuations of fuel consumption during the pandemic. Using the developed model, we project future motor gas consumption for each state. For some states, the gas tax revenues are going to be 10\%-15\% lower than the pre-pandemic level for at least one or two years. 
\end{abstract}

% keywords can be removed
\keywords{COVID-19, Pandemic, Transportation Funding, Infrastructure Management, Resilience}

\section{Introduction}
Transportation spending is considered as one of the largest state expenditures at \$172 billion or 8.1 percent of all state spending in 2019. In the United States, an important part of transportation infrastructure funding sources is considered following the user-pay principle, where an individual pays for the amount he/she uses the transportation system \cite{nga2020transportation, FacklerNiemeier2014UserPays,
HanleyKuhl2011MBUF,
NelsonRowangould2024RuralEquity,
RahmanJuBurris2024VMTFee,
DuncanGraham2013RoadUserFees,
DuncanEtAl2014BenefitTaxation,
DuncanEtAl2014BumpyDesigns,
DuncanEtAl2017TransportPolicyRUC,
KonstantinouLabiGkritza2023EVRevenue,
GreeneBaker2011RUC}. The fees and taxes that consist of most of the transportation funding include:

\begin{itemize}
    \item \textbf{Motor fuel tax.} It includes gasoline and diesel tax. The motor fuel tax has two components, the federal motor fuel tax, and the state motor fuel taxes. For example, in Texas, the state government collects gasoline and diesel fuel tax at \$0.20 per gallon while the federal government collects another \$0.184 per gallon for gasoline and \$0.244 per gallon for diesel. Motor fuel taxes are usually representing around 40\% of all transportation revenue \cite{aashto2016governance}. Among the 50 states in the U.S., 20 of them adjust the fuel tax rate based on inflation or other indexes \cite{urban2017motorfuel}. Some states have begun testing using mileage-based fees to replace the motor fuel tax. Table \ref{tab:fuel_tax_2020} summarizes fuel tax rates for 50 U.S. States and Washington D.C.
\end{itemize}

\begin{table}[h]
    \centering
    \caption{Tax Rate (cent/gallon) of Motor Fuel as of July 2020 \cite{fhwa2020motor}} 
    \label{tab:fuel_tax_2020}
    \vspace{1em}
    \small % Slightly smaller font to ensure a compact look
    \begin{tabular}{lrr p{0.5cm} lrr}
        \toprule
        \textbf{State} & \textbf{Gasoline} & \textbf{Diesel} & & \textbf{State} & \textbf{Gasoline} & \textbf{Diesel} \\
        \cmidrule{1-3} \cmidrule{5-7}
        Alabama        & 26.0 & 27.0 & & Montana        & 32.8 & 30.2 \\
        Alaska         & 8.0  & 8.0  & & Nebraska       & 34.1 & 34.1 \\
        Arizona        & 18.0 & 26.0 & & Nevada         & 23.8 & 27.0 \\
        Arkansas       & 24.8 & 28.8 & & New Hampshire  & 23.8 & 23.8 \\
        California     & 50.5 & 38.5 & & New Jersey     & 37.1 & 40.1 \\
        Colorado       & 22.0 & 20.5 & & New Mexico     & 17.0 & 21.0 \\
        Connecticut    & 25.0 & 46.5 & & New York       & 25.5 & 23.7 \\
        Delaware       & 23.0 & 22.0 & & North Carolina & 36.4 & 36.4 \\
        DC             & 23.5 & 23.5 & & North Dakota   & 23.0 & 23.0 \\
        Florida        & 37.8 & 37.8 & & Ohio           & 38.5 & 47.0 \\
        Georgia        & 27.9 & 31.3 & & Oklahoma       & 20.0 & 20.0 \\
        Hawaii         & 16.0 & 16.0 & & Oregon         & 36.0 & 36.0 \\
        Idaho          & 33.0 & 33.0 & & Pennsylvania   & 57.6 & 74.1 \\
        Illinois       & 39.8 & 47.3 & & Rhode Island   & 35.0 & 35.0 \\
        Indiana        & 32.0 & 52.0 & & South Carolina & 24.0 & 24.0 \\
        Iowa           & 31.0 & 33.5 & & South Dakota   & 30.0 & 30.0 \\
        Kansas         & 24.0 & 26.0 & & Tennessee      & 26.0 & 27.0 \\
        Kentucky       & 24.6 & 21.6 & & Texas          & 20.0 & 20.0 \\
        Louisiana      & 20.0 & 20.0 & & Utah           & 30.0 & 30.0 \\
        Maine          & 30.0 & 31.2 & & Vermont        & 30.5 & 31.0 \\
        Maryland       & 36.3 & 37.1 & & Virginia       & 16.2 & 20.2 \\
        Massachusetts  & 24.0 & 24.0 & & Washington     & 49.4 & 49.4 \\
        Michigan       & 26.3 & 26.3 & & West Virginia  & 35.7 & 35.7 \\
        Minnesota      & 28.5 & 28.5 & & Wisconsin      & 30.9 & 30.9 \\
        Mississippi    & 18.4 & 18.4 & & Wyoming        & 24.0 & 24.0 \\
        Missouri       & 17.0 & 17.0 & & \textbf{Federal Tax} & \textbf{18.4} & \textbf{24.4} \\
        \bottomrule
    \end{tabular}
\end{table}

\begin{itemize}
    \item \textbf{Sales or Other Taxes.} Some states also collect transportation revenue from sales taxes on alcohol and tobacco.
    
    \item \textbf{Vehicle Registration Fees and Taxes}. State charges vehicle license and registration fees, which represent around 20\% of total transportation revenue. These fees include license plate fees, title transfer fees, lien recording fees, and emissions/inspection fees.
     
    \item \textbf{Tolls and congestion pricing}. It is reported that 27 states operate tolls on roads or bridges. Tolls are usually allocated to the maintenance or repayment of some specific roads and only \$1.5 billion toll revenue went to state transportation funds.
    
    \item \textbf{General Funds}. State governments also transfer money from the general fund to the transportation fund. General fund transfers represent 6\% of state highway funding \cite{nga2020transportation}.
\end{itemize}

However, the current transportation funding mechanism has been criticized as not sustainable in the long term as vehicles become more fuel efficient. Gaps in the Highway Trust Fund has to be covered by transferring money from other funds since 2008 \cite{plautz2020gas}.

\section{COVID-19 IMPACT}
\label{sec:headings}

\begin{figure}
    \centering
    \includegraphics[width=0.5\linewidth]{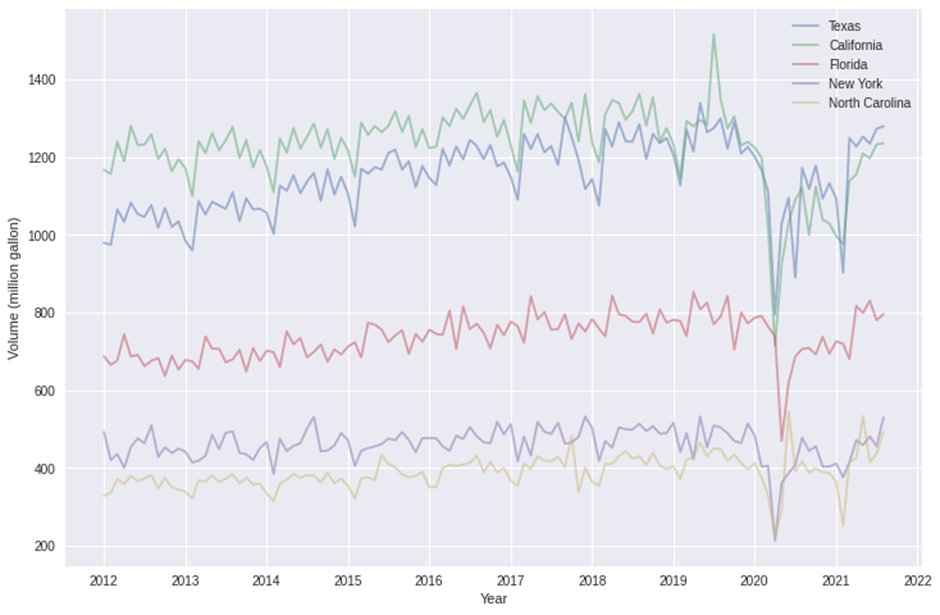}
    \caption{Top Five States with the Largest Gasoline Consumption from January 2012 to August 2021 }
    \label{fig:fig1}
\end{figure}

In 2020, motor fuel consumption was significantly affected because of the COVID-19 pandemic \cite{lamb2020states}. Increasing the use of telecommunication technologies and disruptions to travel and routine activities have all reduced driving \cite{du2020toxic,gao2021prediction}. It was estimated that there were one billion fewer miles each day (or 11.4\% reduction) in April 2020 as compared to the same month in 2019 \cite{olin2020traffic}. American Association of State Highway and Transportation Officials (AASHTO) projected that there will be \$16 billion in revenue reduction in 2020 in the United States and \$37 billion in revenue losses in the next five years \cite{aashto2020covid}. Budget reduction and uncertainty will result in transportation infrastructure condition fluctuations \cite{gao2013augmented,gao2024considering,dhatrak2020considering,gao2019impacts,gao2007using,gao2008robust,gao2010network}. Figure \ref{fig:fig1} shows the top five states’ monthly gasoline consumptions, indicating the consistent and continuous fuel consumption drop starting at the end of 2019 till April and fluctuating after that.

Some states have already taken actions to help counterbalance revenue losses due to COVID-19 \cite{gordon2020state,auerbach2020implications,clemens2020aeiupdate,haughwout2021mlf,bernhardt2021munidebt,clemens2022stimulus,clemens2025health,gao2021fiscalconditions,seegert2023salestax,pew2023pandemicaid}. For example, New Jersey increased the gas tax rate by 9.3 cents per gallon in August to offset the revenue losses \cite{lamb2020states}. Maryland planned to cut \$3 billion in the next 6 years \cite{plautz2020gas}. Maine approved a \$105 million transportation bond in July to compensate for the estimated \$60 million revenue losses in the next two years.

\section{GASOLINE AND FUEL CONSUMPTION DATA}

\begin{figure}
    \centering
    \includegraphics[width=0.5\linewidth]{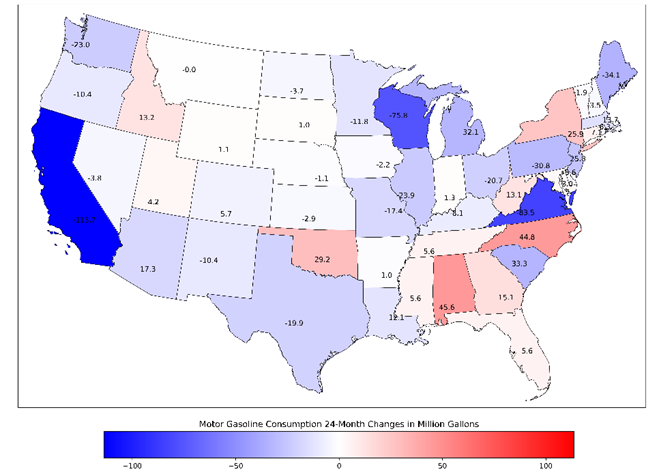}
    \caption{Motor Gasoline Consumption 24-Month Changes in Million Gallons (August 2019 vs. August 2021)}
    \label{fig:fig2}
\end{figure}

\begin{figure}
    \centering
    \includegraphics[width=0.5\linewidth]{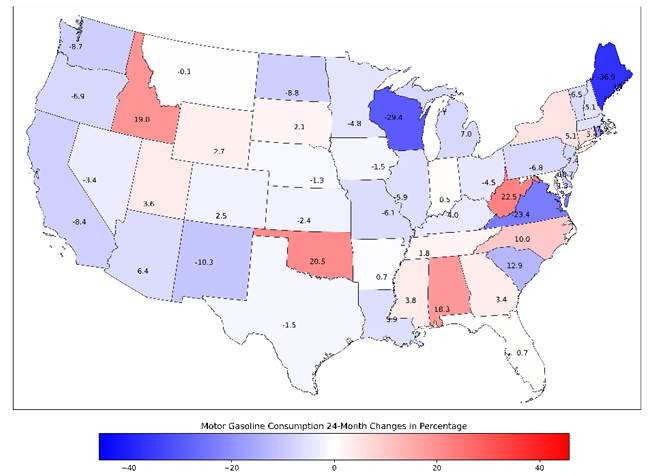}
    \caption{Motor Gasoline Consumption 24-month Changes in Percentage (August 2019 vs. August 2021)}
    \label{fig:fig3}
\end{figure}

In this study, monthly motor fuel consumption of the 50 U.S. states were collected from January 2012 to August 2021. Figures \ref{fig:fig2} and \ref{fig:fig3} show the 24-month change of motor gasoline consumption (August 2019 vs. July 2020) of all the 50 states in million gallons and percentage, respectively. The top two states with the largest declines in gasoline consumption are California (113.7 million gallons less than 2019) and Virginia (83.5 million gallons less than 2019), while top two states with the largest increase in gasoline consumption are Alabama (45.6 million gallons more than 2019), North Carolina (44.8 million gallons more than 2019).

\begin{figure}
    \centering
    \includegraphics[width=0.5\linewidth]{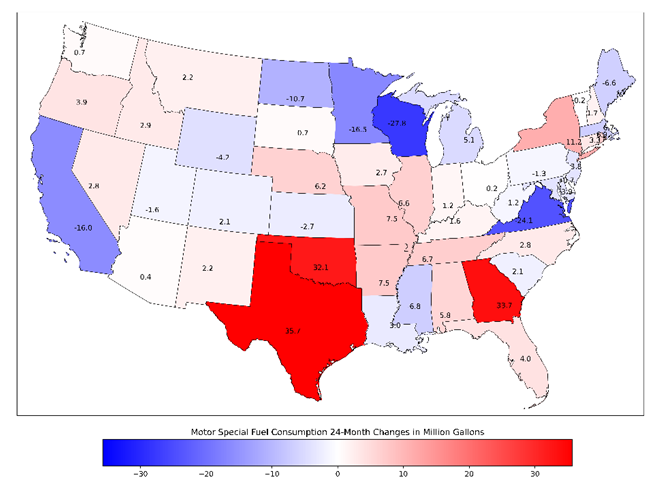}
    \caption{Motor Special Fuel Consumption 24-month Changes in Million Gallons (August 2019 vs. August 2021)}
    \label{fig:fig4}
\end{figure}

\begin{figure}
    \centering
    \includegraphics[width=0.5\linewidth]{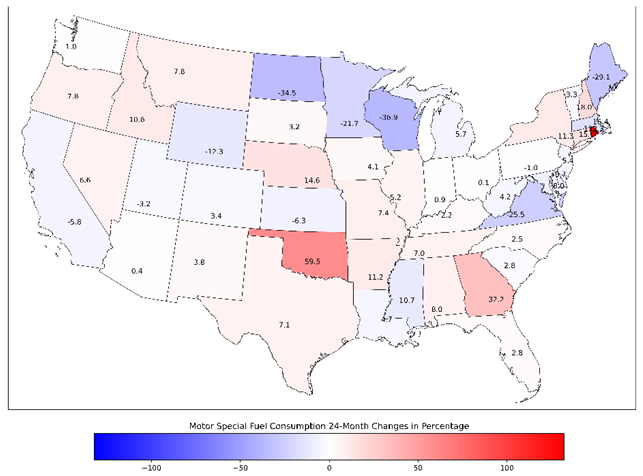}
    \caption{Motor Special Fuel Consumption 24-month Change in Percentage (August 2019 vs. August 2021)}
    \label{fig:fig5}
\end{figure}

\begin{figure}
    \centering
    \includegraphics[width=0.5\linewidth]{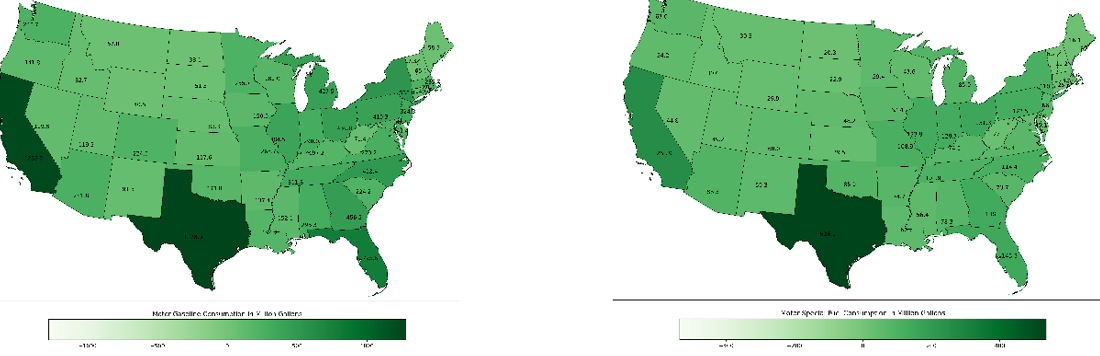}
    \caption{Motor Fuel Consumption (August 2021)}
    \label{fig:fig6}
\end{figure}

Figures \ref{fig:fig4} and \ref{fig:fig5} show the 12-month change of motor special fuel consumption (August 2019 vs August 2021) of all the 50 states in million gallons and percentage, respectively. Special fuel includes primarily diesel. The top two states with the largest declines in special fuel consumption are Illinois (27.8 million gallons less than 2019) and Virginia (24.1 million gallons less than 2019). The top two states with the largest increase in special fuel consumption are Texas (35.7 million gallons more than 2019) and Georgia (33.7 million gallons more than 2019). Figure \ref{fig:fig6} shows the motor fuel consumptions (August 2021) of the 50 states in the U.S.

\section{FUEL CONSUMPTION MODEL}

In this study, machine learning models were developed to predict the motor fuel consumption of each state. Table \ref{tab:tab2} shows the list of influencing variables used in this study, including COVID-19 development, fuel consumptions, and demographic data.

\begin{table}[htbp]
    \centering
    \caption{Data Description}
    \label{tab:tab2}
    \vspace{0.5em}
    \begin{tabularx}{\textwidth}{l X} % "X" column will fill the remaining width
        \toprule
        \textbf{\#} & \textbf{Variables} \\
        \midrule
        1 & Monthly gasoline consumption of the state \\
        2 & Gasoline tax rate of the state \\
        3 & Diesel tax rate of the state \\
        4 & Indicator variables for the month of a year \\
        5 & Indicator variables for state \\
        6 & Population of the state in different years \\
        7 & Number of months since the initial lockdown in April 2020 \\
        \bottomrule
    \end{tabularx}
\end{table}

The 5,800 data points were split into two sets: 70 percent for training and 30 percent for testing. Four machine learning models, including decision tree regression, linear regression, neural network, and random forest regression were trained using the data and the best results of the machine learning methods are shown in Table \ref{tab:tab3}. As one can see in Table \ref{tab:tab3}, decision tree regression produces the best results with a R2-score of 0.989.

\begin{table}[htbp]
    \centering
    \caption{Models' Performance}
    \label{tab:tab3}
    \vspace{0.5em}
    \begin{tabular}{lc} % l: 左对齐模型名, c: 居中对齐得分
        \toprule
        \textbf{Model} & \textbf{$R^2$-Score} \\
        \midrule
        Decision Tree Regression  & 0.979 \\
        Linear Regression         & 0.986 \\
        Neural Network Regression & 0.936 \\
        \textbf{Random Forest Regression} & \textbf{0.989} \\ % 加粗表现最好的模型
        \bottomrule
    \end{tabular}
\end{table}

\begin{figure}[htbp]
    \centering
    \includegraphics[width=0.75\linewidth]{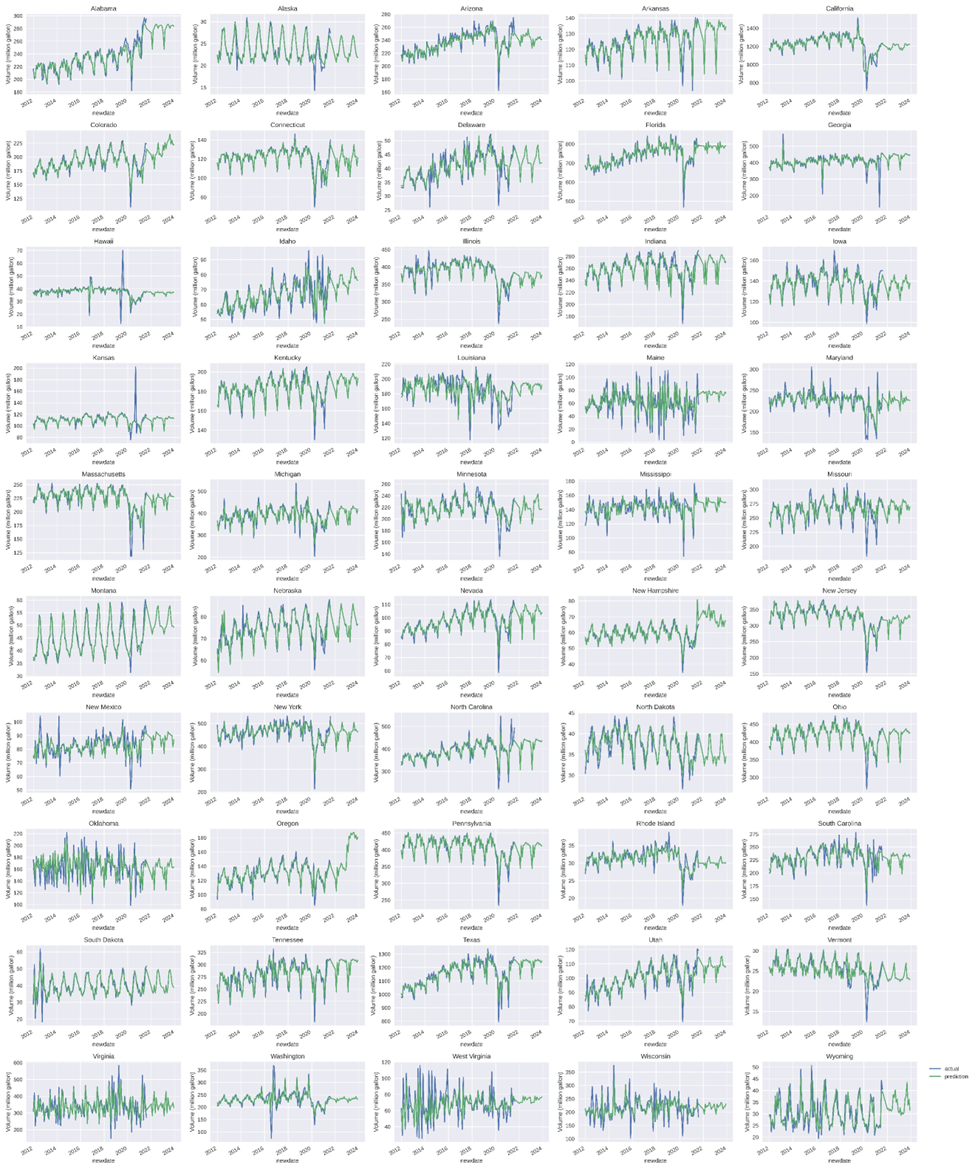}
    \caption{Actual vs. Predicted Gasoline Consumption for the Selected States }
    \label{fig:fig7}
\end{figure}

Figure \ref{fig:fig7} shows the actual gasoline consumption of 50 states during the time period from January 2012 to August 2021. It also includes predicted gasoline volume from January 2012 to December 2024 using the random forest regression model results. As shown in Figure \ref{fig:fig7}, the developed model is able to capture not only the seasonal variations of gasoline consumption but also the long-term trend of the fuel usage. Moreover, the developed model accurately describes the effect of COVID-19 on motor fuel consumption starting from February this year, especially during the national wide lockdown.

\section{Conclusions}

In this paper, machine learning models were developed to predict the motor fuel monthly consumption for each state in the United States. This model can be used to evaluate the impact of COVID-19 on gasoline consumption. As motor fuel tax revenue is a significant part of the transportation infrastructure funding sources, the reduced travel behaviour, and fuel consumption due to COVID-19 result in losses in transportation funding for nearly every state. The model developed in this study can provide stakeholders insights into the influencing factors of motor fuel consumption and different scenarios of future fuel usage. However, as the pandemic is still developing on an unpredictable path in the United States, this model to evaluate the impacts on transportation infrastructure funds will continue to be improved when more data becomes available.

%Bibliography
\bibliographystyle{unsrt}  
\bibliography{references}

@techreport{aashto2016governance,
  author = {{American Association of State Highway and Transportation Officials (AASHTO)}},
  title = {Transportation Governance and Finance: A 50-State Review of State Legislatures and Departments of Transportation},
  year = {2016},
  url = {http://www.financingtransportation.org/pdf/50_state_review_nov16.pdf},
  note = {Consulted 13 December, 2020}
}

@techreport{aashto2020covid,
  author = {{American Association of State Highway and Transportation Officials (AASHTO)}},
  title = {AASHTO Letter to Congress on COVID Phase 4},
  year = {2020},
  month = {July},
  url = {https://policy.transportation.org/wpcontent/uploads/sites/59/2020/07/2020-07-20-AASHTO-Letter-to-Congress-on-COVID-19-Phase-4-FINAL.pdf},
  note = {Consulted 21 December, 2020}
}

@article{du2020toxic,
  author = {Du, J. and Wang, H. and Qiao, F.},
  title = {Transportation-related toxic emissions influenced by public reactions to the COVID-19 pandemic},
  journal = {Journal of Environmental Toxicology Studies},
  volume = {4},
  number = {1},
  pages = {1--4},
  year = {2020},
  doi = {10.16966/2576-6430.128}
}

@misc{fhwa2020motor,
  author = {{Federal Highway Administration (FHWA)}},
  title = {Motor Fuel \& Highway Trust Fund},
  year = {2020},
  url = {https://www.fhwa.dot.gov/policyinformation/motorfuelhwy_trustfund.cfm},
  note = {Consulted 27 December, 2020}
}

@inproceedings{gao2021prediction,
  author = {Gao, L. and Qiao, F. and Li, J. and Lu, P. and Ren, Y. and Zhang, Y.},
  title = {Prediction of the Impact of COVID-19 on US K-12 School Trip Travel Demand},
  booktitle = {Proceedings of the ASCE International Conference on Transportation \& Development (ICTD 2021)},
  year = {2021},
  month = {June},
  note = {Virtual Event, June 8-10, 2021}
}

@article{gao2013augmented,
  author = {Gao, Lu and Guo, Runhua and Zhang, Zhanmin},
  title = {An augmented Lagrangian decomposition approach for infrastructure maintenance and rehabilitation decisions under budget uncertainty},
  journal = {Structure and Infrastructure Engineering},
  volume = {9},
  number = {5},
  pages = {448--457},
  year = {2013},
  publisher = {Taylor \& Francis}
}

@online{lamb2020states,
  author = {Lamb, E.},
  title = {States take action as COVID-19 chokes revenue streams},
  year = {2020},
  url = {https://www.ttnews.com/articles/states-take-action-covid-19-chokes-revenue-streams},
  note = {Consulted 27 December, 2020}
}

@online{olin2020traffic,
  author = {Olin, A.},
  title = {Traffic dropped 66\%, but it came back and the coronavirus followed},
  publisher = {Rice University, Kinder Institute for Urban Research},
  year = {2020},
  month = {July},
  url = {https://kinder.rice.edu/urbanedge/2020/07/27/transportation-traffic-dropped-66-it-came-back-and-covid-19-followed},
  note = {Consulted 13 December, 2020}
}

@online{plautz2020gas,
  author = {Plautz, J.},
  title = {The Gas Tax was Already Broken. The Pandemic could End It},
  year = {2020},
  url = {https://www.smartcitiesdive.com/news/the-gas-tax-was-already-broken-the-pandemic-could-end-it/587653/},
  note = {Consulted 27 December, 2020}
}

@techreport{nga2020transportation,
  author = {{National Governors Association (NGA)}},
  title = {Transportation Funding and Financing during COVID-19},
  year = {2020},
  url = {https://www.nga.org/wp-content/uploads/2020/12/COVID-Memo-Transportation-Funding.pdf},
  note = {Consulted 28 December, 2020}
}

@misc{urban2017motorfuel,
  author = {{Urban Institute}},
  title = {State and Local Finance Initiatives: Motor Fuel Taxes},
  year = {2017},
  url = {https://www.urban.org/policycenters/cross-center-initiatives/state-and-local-finance-initiative/state-and-localbackgrounders/motor-fuel-taxes},
  note = {Consulted 27 December, 2020}
}

@article{FacklerNiemeier2014UserPays,
  title        = {Modern Transportation Funding and User-Pays Principle: Are Drivers Paying for What They Get and Getting What They Pay For?},
  author       = {Fackler, Aaron W. and Niemeier, Debbie},
  journal      = {Transportation Research Record: Journal of the Transportation Research Board},
  year         = {2014},
  volume       = {2450},
  pages        = {1--6},
  doi          = {10.3141/2450-01}
}

@article{HanleyKuhl2011MBUF,
  title        = {National Evaluation of Mileage-Based Charges for Drivers: Initial Results},
  author       = {Hanley, Paul F. and Kuhl, Jon G.},
  journal      = {Transportation Research Record: Journal of the Transportation Research Board},
  year         = {2011},
  volume       = {2221},
  pages        = {10--18},
  doi          = {10.3141/2221-02}
}

@article{NelsonRowangould2024RuralEquity,
  title        = {Data-Driven Analysis of Rural Equity and Cost Concerns for Mileage-Based User Fees in Vermont},
  author       = {Nelson, Clare and Rowangould, Gregory},
  journal      = {Transportation Research Record: Journal of the Transportation Research Board},
  year         = {2024},
  doi          = {10.1177/03611981231206167}
}

@article{RahmanJuBurris2024VMTFee,
  title        = {Income and Geographic Distribution of a Vehicle Miles Traveled Fee: Case Study of Four States},
  author       = {Rahman, Musfira and Ju, HyeMin and Burris, Mark},
  journal      = {Transportation Research Record: Journal of the Transportation Research Board},
  year         = {2024},
  doi          = {10.1177/03611981241295714}
}

@article{DuncanGraham2013RoadUserFees,
  title        = {Road User Fees Instead of Fuel Taxes: The Quest for Political Acceptability},
  author       = {Duncan, Denvil and Graham, John D.},
  journal      = {Public Administration Review},
  year         = {2013},
  volume       = {73},
  number       = {3},
  pages        = {415--426},
  doi          = {10.1111/puar.12045}
}

@article{DuncanEtAl2014BenefitTaxation,
  title        = {Demand for Benefit Taxation: Evidence From Public Opinion on Road Financing},
  author       = {Duncan, Denvil and Graham, John and Nadella, Venkata and Bowers, Ashley and Giroux, Stacey},
  journal      = {Public Budgeting \& Finance},
  year         = {2014},
  volume       = {34},
  number       = {4},
  pages        = {120--142},
  doi          = {10.1111/pbaf.12046}
}

@article{DuncanEtAl2014BumpyDesigns,
  title        = {Bumpy Designs: Impact of Privacy and Technology Costs on Support for Road Mileage User Fees},
  author       = {Duncan, Denvil and Nadella, Venkata and Bowers, Ashley and Giroux, Stacey and Graham, John D.},
  journal      = {National Tax Journal},
  year         = {2014},
  volume       = {67},
  number       = {3},
  pages        = {505--530},
  doi          = {10.17310/ntj.2014.3.01}
}

@article{DuncanEtAl2017TransportPolicyRUC,
  title        = {The road mileage user-fee: Level, intensity, and predictors of public support},
  author       = {Duncan, Denvil and Nadella, Venkata and Giroux, Stacey and Bowers, Ashley and Graham, John D.},
  journal      = {Transport Policy},
  year         = {2017},
  volume       = {53},
  pages        = {70--78},
  doi          = {10.1016/j.tranpol.2016.09.002}
}

@article{KonstantinouLabiGkritza2023EVRevenue,
  title        = {Assessing highway revenue impacts of electric vehicles using a case study},
  author       = {Konstantinou, Theodora and Labi, Samuel and Gkritza, Konstantina},
  journal      = {Research in Transportation Economics},
  year         = {2023},
  volume       = {100},
  pages        = {1--15},
  doi          = {10.1016/j.retrec.2022.101248}
}

@article{GreeneBaker2011RUC,
  title        = {The future of transportation funding: Defining the end-game},
  author       = {Greene, David L. and Baker, Howard H.},
  journal      = {Transportation Research Part D: Transport and Environment},
  year         = {2011},
  volume       = {16},
  number       = {1},
  pages        = {63--71},
  doi          = {10.1016/j.trd.2011.05.003}
}

@article{gordon2020state,
  title   = {State and Local Government Finances in the COVID-19 Era},
  author  = {Gordon, Tracy and Dadayan, Lucy and Rueben, Kim},
  journal = {National Tax Journal},
  year    = {2020},
  volume  = {73},
  number  = {3},
  pages   = {733--758},
  doi     = {10.17310/ntj.2020.3.05}
}

@techreport{auerbach2020implications,
  title       = {Implications of the Covid-19 Pandemic for State Government Tax Revenues},
  author      = {Auerbach, Alan J. and Gale, William G. and Lutz, Byron F. and Sheiner, Louise},
  institution = {National Bureau of Economic Research},
  type        = {NBER Working Paper},
  number      = {27426},
  year        = {2020},
  doi         = {10.3386/w27426},
  url         = {https://www.nber.org/papers/w27426}
}

@techreport{clemens2020aeiupdate,
  title       = {The COVID-19 Pandemic and the Revenues of State and Local Governments: An Update},
  author      = {Clemens, Jeffrey and Veuger, Stan},
  institution = {American Enterprise Institute},
  year        = {2020},
  month       = sep,
  url         = {https://aei.org/wp-content/uploads/2020/09/The-COVID-19-Pandemic-and-the-Revenues-of-State-and-Local-Governments.pdf}
}

@techreport{haughwout2021mlf,
  title       = {The Municipal Liquidity Facility},
  author      = {Haughwout, Andrew and Hyman, Ben and Shachar, Or},
  institution = {Federal Reserve Bank of New York},
  type        = {Staff Report},
  number      = {985},
  year        = {2021},
  month       = sep,
  url         = {https://www.newyorkfed.org/medialibrary/media/research/staff_reports/sr985.pdf}
}

@techreport{bernhardt2021munidebt,
  title       = {The Impact of Covid-19 Related Policy Responses on Municipal Debt Markets},
  author      = {Bernhardt, Robert and D'Amico, Stefania and Sordo Palacios, Santiago I.},
  institution = {Federal Reserve Bank of Chicago},
  type        = {Working Paper},
  number      = {2021-14},
  year        = {2021},
  month       = sep,
  doi         = {10.21033/wp-2021-14},
  url         = {https://www.chicagofed.org/publications/working-papers/2021/2021-14}
}

@techreport{clemens2022stimulus,
  title       = {Was Pandemic Fiscal Relief Effective Fiscal Stimulus? Evidence from Aid to State and Local Governments},
  author      = {Clemens, Jeffrey and Hoxie, Philip G. and Veuger, Stan},
  institution = {National Bureau of Economic Research},
  type        = {NBER Working Paper},
  number      = {30168},
  year        = {2022},
  month       = jun,
  doi         = {10.3386/w30168},
  url         = {https://www.nber.org/papers/w30168}
}

@techreport{clemens2025health,
  title       = {Health Impacts of Federal Pandemic Aid to State and Local Governments},
  author      = {Clemens, Jeffrey and Mahajan, Anwita},
  institution = {National Bureau of Economic Research},
  type        = {NBER Working Paper},
  number      = {33699},
  year        = {2025},
  month       = apr,
  doi         = {10.3386/w33699},
  url         = {https://www.nber.org/papers/w33699}
}

@techreport{gao2021fiscalconditions,
  title       = {State and Local Governments: Fiscal Conditions During the COVID-19 Pandemic in Selected States},
  author      = {{U.S. Government Accountability Office}},
  institution = {U.S. Government Accountability Office},
  type        = {Report to Congressional Committees},
  number      = {GAO-21-562},
  year        = {2021},
  month       = jul,
  url         = {https://www.gao.gov/assets/gao-21-562.pdf}
}

@article{seegert2023salestax,
  title   = {The COVID-19 state sales tax windfall},
  author  = {Seegert, Nathan and Dean, Phil and Gaulin, Maclean and Yang, Mu-Jeung},
  journal = {International Tax and Public Finance},
  year    = {2023},
  doi     = {10.1007/s10797-023-09778-w},
  url     = {https://link.springer.com/article/10.1007/s10797-023-09778-w}
}

@techreport{pew2023pandemicaid,
  title       = {Pandemic Aid: How States Safeguarded Against Future Budget Challenges},
  author      = {{The Pew Charitable Trusts}},
  institution = {The Pew Charitable Trusts},
  year        = {2023},
  month       = dec,
  url         = {https://www.pew.org/en/research-and-analysis/reports/2023/12/pandemic-aid-how-states-safeguarded-against-future-budget-challenges},
  note        = {Updated February 28, 2024}
}

@article{gao2008robust,
  title={Robust optimization for managing pavement maintenance and rehabilitation},
  author={Gao, Lu and Zhang, Zhanmin},
  journal={Transportation research record},
  volume={2084},
  number={1},
  pages={55--61},
  year={2008},
  publisher={SAGE Publications Sage CA: Los Angeles, CA}
}

@techreport{gao2010network,
  title={Network-level multi-objective optimal maintenance and rehabilitation scheduling},
  author={Gao, Lu and Xie, Chi and Zhang, Zhanmin},
  year={2010}
}

@techreport{gao2007using,
  title={Using Markov process and method of moments for optimizing management strategies of pavement infrastructure},
  author={Gao, Lu and Zhang, Zhanmin and Tighe, Susan Louise},
  year={2007}
}

@article{gao2019impacts,
  title={Impacts of seasonal and annual weather variations on network-level pavement performance},
  author={Gao, Lu and Hong, Feng and Ren, Yi-Hao},
  journal={Infrastructures},
  volume={4},
  number={2},
  pages={27},
  year={2019},
  publisher={MDPI}
}

@article{dhatrak2020considering,
  title={Considering deterioration propagation in transportation infrastructure maintenance planning},
  author={Dhatrak, Omkar and Vemuri, Venkata and Gao, Lu},
  journal={Journal of Traffic and Transportation Engineering (English Edition)},
  volume={7},
  number={4},
  pages={520--528},
  year={2020},
  publisher={Elsevier}
}

@article{gao2024considering,
  title={Considering the spatial structure of the road network in pavement deterioration modeling},
  author={Gao, Lu and Yu, Ke and Lu, Pan},
  journal={Transportation Research Record},
  volume={2678},
  number={5},
  pages={153--161},
  year={2024},
  publisher={SAGE Publications Sage CA: Los Angeles, CA}
}

\end{document}